# PERFORMANCE OF PIP-II HIGH-BETA 650 CRYOMODULE AFTER TRANSATLANTIC SHIPPING*

J. Ozelis[†], M. Barba, J. Bernardini, C. Contreras-Martinez, D. Crawford, J. Dong, V. Grzelak, P. Hanlet, J. Holzbauer, Y. Jia, S. Kazakov, T. Khabiboulline, J. Makara, N. Patel, V. Patel, L. Pei, D. Peterson, Y. Pischalnikov, D. Porwisiak, S. Ranpariya, J. Steimel, N. Solyak, J. Subedi, A. Sukhanov, P. Varghese, T. Wallace, M. White, S. Wijethunga, Y. Xie, S. Yoon,  Fermilab, Batavia, USA

*Abstract*

After shipment to the Daresbury Lab and return to Fermilab, the prototype HB650 cryomodule underwent another phase of 2K RF testing to ascertain any performance issues that may have arisen from the transport of the cryomodule. While measurements taken at room temperature after the conclusion of shipment indicated that there were no negative impacts on cavity alignment, beamline vacuum, or cavity frequency, testing at 2K was required to validate other aspects such as tuner operation, cavity coupling, cryogenic system integrity, and cavity performance. Results of this latest round of limited 2K testing will be presented.

## INTRODUCTION

The PIP-II Superconducting Linac features 23 cryomodules of 5 different types. The highest energy section of the Linac is comprised of 9 Low-Beta (LB) 650 MHz cryomodules with 4 cavities, and 4 High-Beta (HB) cryomodules with 6 cavities. Except for a prototype HB cryomodule, the remaining 650 MHz cryomodules (including the refurbishment of the HB650 prototype cryomodule) are to be assembled by international partners (CEA and INFN for LB, UKRI's Daresbury Lab for HB) as in-kind contributions to the PIP-II project. As a result, 13 of the cryomodules to be used in the P IP-II Linac will require that they undergo transatlantic shipment to the US.

Reducing the risk that this transatlantic shipment results in performance degradation or damage to these cryomodules is paramount. To that end two approaches have been taken to understand the environmental conditions posed by such transport and to mitigate risks:

- Shipment of a fully instrumented "dummy load" to perform preliminary verification of shipment methods and transport fixtures
- Shipment of the prototype HB650 cryomodule (HB650 pCM) to Daresbury Lab after completion of 2K testing at Fermilab's PIP2IT facility, and re-test upon return

Shipment of the dummy load was performed during summer 2022. As reported elsewhere [1], the results indicated that the fixturing and processes (adopted from similar experience shipping LCLS-II cryomodules from FNAL to SLAC) worked extremely well, and minor changes were pursued to incorporate lessons learned.

The shipment of the actual pCM took place late in 2023, after the first round of RF testing of the cryomodule at 2K to characterize initial performance was completed in the PIP2IT facility at Fermilab. The shipment configuration, preparation, and process for shipment of the completed HB650 pCM is described in detail in [2].

## TRANSPORT-CRITICAL CRYOMODULE PERFORMANCE PARAMETERS

A number of critical performance parameters or system/component functionalities were the primary focus of the post-transport study. These include :

- Insulating vacuum integrity
- Beamline vacuum integrity
- Cryogenic circuit integrity
- Cavity alignment preservation
- Tuner operation
- Cryogenic valve actuator operations
- Instrumentation
- High-power coupler Qext
- Cavity performance (gradient, field emission, and quality factor)
- Thermal performance consistency

Each of these areas were subject to test/measurement, and results to be compared with performance observed before shipment.

## TEST APPROACH AND PLAN

Upon return from the UK, the HB650 pCM was partially re-assembled, as the top port and its internal components (e.g., heat exchanger, thermal and magnetic shields) and power coupler air-side components were removed for transport. This re-assembly activity also provided an opportunity to make several modifications to the cryomodule cryogenic systems and mechanical structure, based on results and analysis from earlier thermal studies. A check valve was installed in the 2K relief line, additional MLI was added, and temperature sensors and heaters were installed on the cryogenic valve thermal intercepts, to better characterize heat transfer and examine the possibility of thermo-acoustic oscillations [3].

After re-assembly the cryomodule was transferred for 2K RF testing to the PIP-II Integrated Test facility (PIP2IT) [4][5] (Figure 3)Aft. A test plan was developed in order to systematically evaluate the critical performance parameters that could be affected by long-distance transport, and could be divided into room temperature and cryogenic tests/measurements, as follows :

- Room temperature testing & validation
  - Instrumentation/sensor readout
  - Pumpdown and measure insulating vacuum
  - Measure beamline vacuum
  - Verify cavity alignment
  - Measure cavity frequencies and power coupler $Q_{ext}$

- Cryogenic (2K) testing/validation
  - Instrumentation functionality
  - Beamline vacuum
  - Cryogenic piping/valve integrity
  - Tuner range/operation
  - Coupler thermal performance
  - Cavity performance (gradient, field emission and/or multipacting behaviour, $Q_0$)

During pre-transport testing of the pCM, all 6 cavities were characterized. This however was an arduous process, as it required switching the high power RF distribution system from cavity to cavity, as only one RF amplifier was available. In order to accommodate the shorter schedule for post-shipment testing, only two cavities (5 & 6) were RF tested after transport. And while extensive thermal studies were performed, direct pre-and post-transport comparison is not useful as modifications to the cryomodule in order to reduce static 2K heat load were pursued as mentioned previously.

## RESULTS

Results from post transport testing and evaluations are summarized below.

### Insulating and Beamline Vacuum

Due to the required removal of the coupler air-side components and top port, the cryomodule could not be shipped with the insulating vacuum space under vacuum. Once the module arrived back at FNAL, and all flanges, coupler components etc. re-installed, the cryomodule insulating vacuum space was pumped down and leak checked, and no leaks were found.

The cryomodule was transported with the beamline under vacuum, and a NEG pump was attached to the beamline and actively monitored. Beamline vacuum was 1.1x $10^{-9}$ mbar upon departure from FNAL, and when measured at UKRI (about 4 days later) had actually improved to 8x$10^{-11}$ mbar due to ambient temperature differences (see Figure 1). Upon return to FNAL, the beamline vacuum was 2.0x$10^{-9}$ mbar.

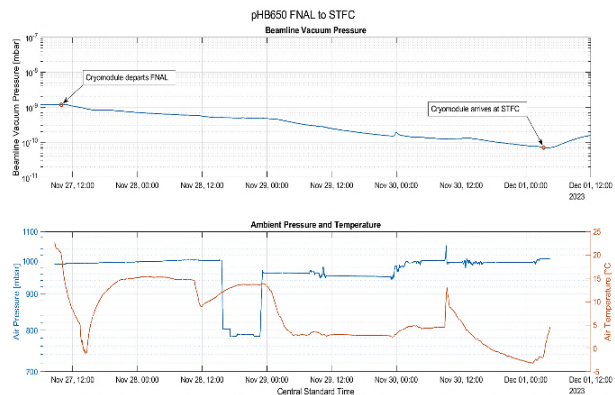

Figure 1. Beamline vacuum during transport from FNAL to UKRI.

### Cavity Alignment

Cavity positions in the cryomodule are measured using the HBCAM system [6]. This system uses permanently mounted high definition cameras that view targets mounted on the cavities through a precision optical viewport. No significant change in cavity positions were observed after round trip transport. Figure 2 shows the displacements of the HBCAM targets before and after transport, at room temperature. The maximum shift is about 190 μm.

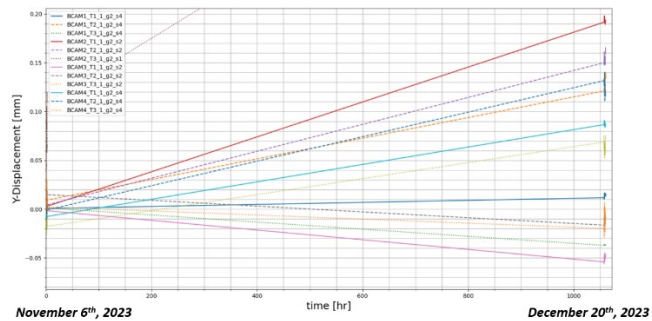

Figure 2. Cavity positions before and after transport, measured at 300K.

### Warm RF measurements

Measurements of cavity frequency and input coupler $Q_{ext}$ were performed while the cryomodule was warm, before and after shipment. Results can be found in Table 1. The largest difference was under 6%, and all differences can be attributed to measurement uncertainty.

Table 1: Measured coupler $Q_{ext}$ before and after shipment

| Cavity | Before shipment | After shipment |
|--------|-----------------|----------------|
| 1 | 8.06 x $10^6$ | 8.17 x $10^6$ |
| 2 | 6.71 x $10^6$ | 6.81 x $10^6$ |
| 3 | 6.59 x $10^6$ | 6.22 x $10^6$ |
| 4 | 1.34 x $10^7$ | 1.35 x $10^7$ |
| 5 | 9.32 x $10^6$ | 9.17 x $10^6$ |
| 6 | 9.09 x $10^6$ | 9.43 x $10^6$ |

### Cavity frequency consistency and tuner performance

After warm measurements were concluded, the module was cooled down to 2K, and no issues with cryogenic operations or leaks in the helium circuits were observed (Figure 3). Once the cavities were at 2K, cavity frequencies in the "cold landing" position (defined as the frequency at which the tuners begin to be disengaged, which prevent inelastic deformation of the cavity during warm-up) were measured and compared with values measured during the previous 2K test run. Results are given in Table 2. Note that cavity 4 did experience a frequency shift, but this was traced to it not being returned to the "cold-landing" frequency (as opposed to the operational frequency of 650.000 MHz) before the previous warmup.

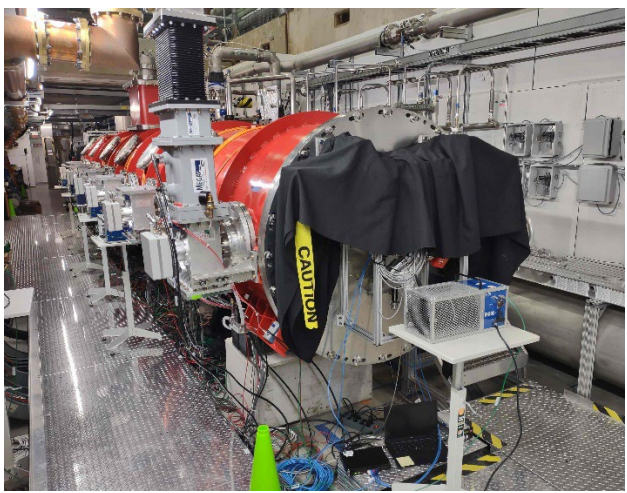

Figure 3: HB650 pCM installed in PIP2IT for post-transport testing at 2K

All cavity tuners were exercised in order to tune the cavities to the operational frequency, and no issues with tuner operations were uncovered (except for a wiring error which was corrected). Tuner range was also examined in select cases and found to still meet requirements (Table 3).

Table 2: Measured cavity frequencies (in MHz) at "cold landing" positions before and after shipment

| Cavity | Before shipment | After shipment |
|---|---|---|
| 1 | 650.0573 | 650.061 |
| 2 | 649.9754 | 649.965 |
| 3 | 649.709 | 649.702 |
| 4 | 650.1164 | 649.963 |
| 5 | 650.0762 | 650.079 |
| 6 | 650.0338 | 650.038 |

Table 3: Select measured cavity tuner ranges

| Cavity | Range (kHz) |
|---|---|
| 1 | 209 |
| 2 | 176 |
| 4 | 166 |
| 5 | 270 |
| 6 | 229 |

### Cavity 2K RF performance

Cavities 5 & 6 were-retested to verify their RF performance after shipping. Particular attention was paid to gradient limit, MP onset and conditioning, FE onset and level, and cavity quality factor ($Q_0$). Additionally, couplers were operated off-resonance to maximum power to determine if there were any change sin thermal, vacuum, or MP performance. Results are summarized in Table 4.

Table 4 : Comparison of relevant cavity performance parameters before and after transport.

| Cavity | Before shipment | After shipment | Before shipment | After shipment |
|---|---|---|---|---|
| | $Q_{ext}$ | | $Q_0$ | |
| 5 | 8.9E+06 | 8.1E+06 | 1.5E+10 | 1.7E+10 |
| 6 | 9.7E+06 | 9.9E+06 | 3.7E+10 | 4.6E+10 |
| | FE Onset (MV/m) | | MP Onset (MV/m) | |
| 5 | 14.5 | 16.5 | 7.0 | 14.0 |
| 6 | 16.2 | 17.0 | 12.8 | 13.0 |
| | Emax (MV/m) | | | |
| 5 | 20.0 | 20.2 | | |
| 6 | 20.5 | 20.7 | | |

From Table 4 it is clear that no significant changes in cavity performance have been observed after the shipment of the cryomodule. There are slight changes (improvement) in FE and MP onset, and these can be attributed to some processing "memory" from the initial round of tests.

### Thermal Performance

As mentioned earlier, direct pre- and post-shipment thermal performance cannot be easily compared as modifications were made to the HB650 pCM specifically to reduce the heat loads observed in earlier testing.

As a result of mitigations and additional testing, most of the sources of the high 2K heat loads have been identified, and an updated analysis indicates that the 2K heat loads for the HB650 production cryomodules are expected to be below 14.5 W with cavities at 4K, and below 19 W with cavities at 2K [3]. Further investigations will be conducted during the disassembly of this cryomodule.

## CONCLUSIONS

The HB650 PCM was transported to the UK and back to FNAL after 2K RF testing in PIP2IT, following protocols established during a successful transport of a dummy load. Upon receipt, the CM was re-assembled, some modifications performed to address higher than expected heat loads, the then module was re-tested at PIP2IT.

No noticeable degradation in cryomodule performance was uncovered during this post-transport testing, all systems and components maintained their original functionality. This is a significant result considering that up to 13 completed cryomodules must be transported from partner laboratories overseas to FNAL, and demonstrates the appropriateness of the engineering, analysis, and planning that went into developing, prototyping, and then implementing the cryomodule transport protocol for PIP-II.